\documentclass{ws-procs975x65}
\usepackage{hyperref}

\def\beq{\begin{equation}}
\def\eeq{\end{equation}}

\begin{document}

\title{Quark-nova compact remnants: \\ Observational signatures in astronomical data and implications to compact stars }
\author{Rachid Ouyed$^*$, Denis Leany, Nico Koning, Zachary Shand}

\address{Physics and Astronomy, University of Calgary,\\
Calgary, Alberta, Canada\\
$^*$E-mail: rouyed@ucalgary.ca}

\begin{abstract}
Quark-novae leave behind quark stars with a surrounding metal-rich fall-back (ring-like) material. These compact remnants have high magnetic 
fields and are misconstrued as magnetars; however, several observational features allow us to distinguish 
a quark star (left behind by a quark-nova) from a neutron star with high magnetic field. In our model, bursting activity 
is expected from intermittent accretion events from the surrounding fall-back debris leading to 
X-ray bursts (in the case of a Keplerian ring) or gamma ray bursts (in the case of a co-rotating shell). 
The details of the spectra are described by a constant  background X-ray luminosity from the expulsion of magnetic flux 
tubes which will be temporarily buried by bursting events caused by accretion of material onto 
the quark star surface. These accretion events emit high energy photons and heat up the 
quark star and surrounding debris leading to hot spots which may be observable as 
distinct blackbodies. Additionally, we explain observed spectral line features as atomic lines from r-process 
material and explain an observed anti-glitch in an AXP as the transfer of angular momentum from a surrounding 
Keplerian disk to the quark star. 
\end{abstract}


\bodymatter


\section{Introduction}
Analogous to a core-collapse supernova, a quark-nova is the explosion of a neutron star which leaves behind a quark star. 
In the quark-nova model, the compact remnant left behind is a quark star composed of deconfined up, down and strange 
quarks in a color-superconducting phase. The quark star has many observable similarities to a neutron star and 
is especially likely to be mislabelled as a magnetar due to its intrinsically high magnetic field at birth. 
As a compact remnant, the quark star distinguishes itself from other compact objects by the properties of its constituent quark matter and
its formation mechanism which ejects an r-process rich crust which leaves behind metal rich debris. These properties 
cause the quark-star to emit in X-ray and gamma ray and causes intermittent bursting phases which 
make it an excellent candidate for not just magnetars, but also soft gamma repeaters (SGRs), anomalous X-ray pulsars (AXPs) 
rotating radio transients (RRATs) and X-ray dim isolated neutron stars (XDINs).

\section{Quark-novae}
Quark-novae convert the parent neutron star into a quark star.\cite{2004A&A...420.1025O} The explosive event ejects an outer layer of the neutron 
star crust ($\sim10^{-3} \textrm{M}_\odot$) and leaves behind a quark star. The relativistic ejecta ($\gamma \geq 10 $) 
is initially an extremely dense ($10^{14}$ g/cc) and neutron rich environment ideal for formation 
of heavy elements through the r-process (see Refs.~\refcite{2007A&A...471..227J}, \refcite{rJava1} and \refcite{rJava2} for more details). 
This ejecta may be visible following the quark-nova either by interaction with surrounding material as 
a super-luminous or double-humped supernova,\cite{SLSNKostka2014,1674-4527-13-12-007}
or by the decay of unstable r-process isotopes as a fast radio burst.\cite{2015arXiv150508147S} Additional observational 
evidence for this r-process rich material may exist in the form of gravitationally bound fall-back material surrounding 
the quark-star. This fall-back material can either form a co-rotating shell, or a Keplerian disk depending on the period of 
the parent neutron star. The large amount of material ($10^{-7} \textrm{M}_\odot$) in these rotating structures provides 
additional emission spectra through both black-body radiation and accretion onto the bare quark star.

\subsection{Quark-nova compact remnant: aligned rotator}
The quark star remnant of the quark-nova is left in a colour-superconducting phase and the 
magnetic field of the quark star becomes quantized and constrained to vortices in an Abrikosov lattice. 
This results a compact object whose surface magnetic field becomes 
much stronger and is forced into alignment with its rotation through the Meissner effect.\citep{2006ApJ...653..558O} 
As the star ages, spin-down causes magnetic field vortex expulsion which leads to heating of the quark star surface and 
X-ray emission from magnetic reconnection events.\cite{2010PhRvD..81d3005N} This ultimately leads to the 
slow decay of the quark star magnetic field and produces a compact body which cannot exhibit the lighthouse 
pulsations of a neutron star, but will instead appear in X-ray. The X-ray emission will decay over time 
and is consistent with observations of XDINS and SGRs (see Fig.~2 in \refcite{2011MNRAS.415.1590O}). The luminosity 
of the vortex band is given by: 
\begin{equation}
\label{eq:xrayLuminosity}
	L_\textrm{X} \approx 2.01 \times 10^{35} \eta_\textrm{X} \dot{P}^2_{-11} \textrm{ ergs}^{-1} 
\end{equation}
where $\eta_\textrm{X}$ is an efficiency factor for conversion of magnetic energy to radiation.\cite{2004A&A...420.1025O}

\section{Quark-nova connection to AXPs and SGRs}
During the quark-nova a small fraction ($\sim10^{-7} \textrm{M}_\odot$) of the ejecta will remain gravitationally bound. 
Depending of the spin of the neutron star progenitor, this debris will form into either a co-rotating shell ($\textrm{P} > 10 $ ms) 
or a Keplerian disk ($\textrm{P} \sim  5$ ms). As with the escaped ejecta, this material will be rich in 
r-process ($\textrm{Z} > 26 $) elements. This debris surrounding the quark star provides an additional source of material 
which may radiate or interact with the quark star. In both cases, it may be possible to identify the presence of two distinct 
blackbodies (where one is the quark star and the second is the surrounding debris) at two different temperatures.

\subsection{Co-rotating debris: SGRs }
\label{sec:sgrs}
For slow rotating neutron star progenitors, the fall-back material will settle into a co-rotating degenerate shell surrounding 
the quark-star,\cite{2007A&A...473..357O}  where the surrounding shell is supported by the quark star's magnetic field pressure. Because 
of this, there is a critical latitude at which material near the poles does not have sufficient magnetic support to 
overcome the quark star's gravitational field. As the star ages, magnetic field decay, contraction of the co-rotating 
shell, hydrodynamic instabilities and changes in the stability point cause accretion of material from the shell onto 
the quark star's surface. These accretion phases cause sudden shearing off of material from the co-rotating shell 
and results in the release gamma rays as the accreted matter from the shell impact onto the surface of the quark star. 
During quiescent phases, these SGRs will be emitting in X-ray from the magnetic field decay as described by Eq.~\ref{eq:xrayLuminosity}.

\subsection{Keplerian debris: AXPs}
Rapidly rotating neutron stars will instead form a degenerate Keplerian disk surrounding the 
quark-star.\cite{2007A&A...475...63O} The magnetic field of the quark star slowly penetrates through the 
degenerate ring and threads the inner ring. This penetration front (Bohm Front) proceeds through the degenerate layers of the ring 
and forces co-rotation of the inner threaded portion of the now non-degenerate inner section of the debris. 
At the front, before complete penetration and forced co-rotation with the magnetic field, the non-degenerate material  
is expelled from the disk through a Kelvin-Helholtz instability and accreted onto the quark star's surface. This leads to a stream of material from the Keplerian 
disk which is accreted onto the 
quark star creating hot-spots near the poles. Dissipation of magnetic bubbles generated at the ring during these outburst provides a source of 
transient radio emission which has been associated with some AXPs following bursting periods. \cite{2010A&A...516A..88O} 
When these accretion events happen the inner ring of the disk will be heated by 
radiation from the surface of the quark star. During quiescent phases, the 
quark star--ring system should have two emission spectra (X-ray and black body). During bursting, however, 
the inner portion of the ring will heat up (increasing its black body temperature) 
and hot spots on the quark star surface will become visible as an additional black body.
 Of course, as the material from the Keplerian ring is transferred onto the quark star, 
the star may be spun up or down depending on radial drift velocity of the accreted material which changes 
with the radial distance. 

As the quark star and Keplerian disk age, the ring begins to spread out and eventually 
succumbs to gradual mass leakage. The gradual accretion from the surrounding ring 
provides a steady stream of matter onto the surface of the quark star leading to 
an accretion dominated phase.\cite{2011MNRAS.415.1590O} Now the quiescent phase 
of the quark star's X-ray luminosity is dominated by the accretion luminosity 
of hot spots on the star. As the debris is accreted the luminosity slowly decreases and, 
once the ring has been completely consumed, the quark star will return 
to the vortex band and emit primarily X-rays from the magnetic reconnection of the 
ever decaying magnetic field (see Fig.~2 of Ref.~\refcite{2011MNRAS.415.1590O}). 
This means that despite their differences early on, AXPs and SGRs will both eventually 
converge to the vortex band and eventually cease bursting activity after they 
have consumed the entirety of the fall-back material from the original quark-nova. 

\section{Quark-nova bursting signatures}
When a quark-star enters into its bursting phases, it will energize the surrounding material. 
This transfer of energy heats up the surrounding debris and may cause atomic spectral line emission 
or cause changes to the rotational mode of the Keplerian disk.

\subsection{Atomic 13 keV lines}
Spectral line features around 13 keV have been detected in several different AXPs. The 
common explanation for these spectral lines has been a cyclotron emission in 
the atmosphere of a magnetar. This explanation has been strained by the repeated observations of 
several different line features at the same frequency in different AXPs. Additionally, the magnetic field strength 
inferred (assuming proton cyclotron emission) is inconsistent with 
the characteristic magnetic field strength (inferred from period and spin-down measurements). 
Instead, we suggest that this observed spectral feature may be an atomic transition line.
Since the fall-back material from the quark-nova ejecta is so neutron rich, the degenerate ring or 
co-rotating shell is composed of r-process elements. Normally, these isotopes are too dim to detect spectroscopically; however, 
during bursting phases these elements are heated up and become spectroscopically visible. The frequencies at 
which these elements emit is then completely independent of magnetic field strength and would be at identical 
between frequencies for all AXPs. Analysis of the atomic line strengths of elements suggests that this 
is a 13 keV emission feature from a strontium or rubidium line.\cite{2013RMxAA..49..351K,2014RMxAA..50..189K} 
The number of observed 13 keV lines makes it unlikely that this is a proton cyclotron line since that would require 
the magnetic fields to be the same in each magnetar. As more of these spectral lines are observed, it may begin 
to statistically rule out magnetars altoghether and atomic line spectra provides a natural explanation for a common
13 keV line to be present in all AXPs.

\subsection{Anti-glitches}
The observation of an anti-glitches associated with AXP 1E2259+586 proved to be 
a puzzling phenomena. In our model, AXP bursts are generated by accretion of material from 
a surrounding disk.\cite{2014Ap&SS.352..715O} These bursting phases re-energize the surrounding material and 
may cause retrograde motion in a surrounding Keplerian disk. For this 
particular AXP, the hypothesis is that the observed burst in 2002 caused a reversal in the inner ring 
of the Keplerian disk. Then, in 2012, material from the ring was accreted causing spin-down due to the transfer of 
angular momentum from the now retrograde inner ring. As a result, further bursts from the same AXP may cause 
further anti-glitches if the 2002 burst was energetic enough to reverse several layers of the surrounding inner ring. 
Coincident observation of atomic r-process spectral features would further strengthen this hypothesis. 

\section{Conclusion}
The quark-nova model proposes that quark stars are the compact remnant left behind 
following the explosion of a neutron star. The explosion process leaves behind debris 
which form a gravitationally bound disk or ring around the quark star. This quark star--debris 
system emits in X-ray and is subject to several bursting phases over the course of its 
lifetime leading to either SGRs, or AXPs. The X-ray luminosity generated by vortex expulsion is consistent 
with observations of SGRs and AXPs and suggests a common ancestry for both. This picture is similar to magnetar models for AXPs and 
SGRs, but can be distinguished by detailed analysis of the observed spectra. 
Based on our model we expect that future bursting events 
should be associated with both low and high B magnetars and may be coincident with anti-glitches and r-proceess atomic spectral lines.

\section*{Acknowledgments}

This work is funded by the Natural Sciences and
Engineering Research Council of Canada.

\bibliographystyle{ws-procs975x65}
\bibliography{main}

\end{document}